\documentclass[%
 aip,
 sd,%
 amsmath,amssymb,
 reprint,%
]{revtex4-1}

\usepackage{graphicx}% Include figure files
\usepackage{dcolumn}% Align table columns on decimal point
\usepackage{bm}% bold math
\usepackage{color} 
\usepackage{floatrow}
\usepackage{textcomp}

\begin{document}

%\preprint{AIP/123-QED}

\title{Experimental set-up for particle detection in solid crystals of inert gasses }

\author{M. Guarise}
 \email{marco.guarise@lnl.infn.it}
\affiliation{Dipartimento di Fisica e Scienze della Terra, Via G. Saragat 1, 44122 Ferrara, Italy }
\affiliation{INFN Laboratori Nazionali Legnaro, Viale dell'Universit\`a 2, 35020 Legnaro (Pd), Italy }
\author{C. Braggio}
\affiliation{Dipartimento di Fisica e Astronomia and INFN Sezione di Padova, Via F. Marzolo 8, I-35131 Padova, Italy }
\author{R. Calabrese}
\affiliation{Dipartimento di Fisica e Scienze della Terra, Via G. Saragat 1, 44122 Ferrara, Italy }
\affiliation{INFN Sezione di Ferrara, Via G. Saragat 1, 44122 Ferrara, Italy}
\author{G. Carugno}
\affiliation{Dipartimento di Fisica e Astronomia and INFN Sezione di Padova, Via F. Marzolo 8, I-35131 Padova, Italy }
\author{A. Dainelli}
\affiliation{INFN Laboratori Nazionali Legnaro, Viale dell'Universit\`a 2, 35020 Legnaro (Pd), Italy }
\author{A. Khanbekyan}
\affiliation{Dipartimento di Fisica e Scienze della Terra, Via G. Saragat 1, 44122 Ferrara, Italy }
\affiliation{INFN Sezione di Ferrara, Via G. Saragat 1, 44122 Ferrara, Italy}
\author{E. Luppi}
\affiliation{Dipartimento di Fisica e Scienze della Terra, Via G. Saragat 1, 44122 Ferrara, Italy }
\affiliation{INFN Sezione di Ferrara, Via G. Saragat 1, 44122 Ferrara, Italy}
\author{M. Poggi}
\affiliation{INFN Laboratori Nazionali Legnaro, Viale dell'Universit\`a 2, 35020 Legnaro (Pd), Italy }
\author{L. Tomassetti}
\affiliation{Dipartimento di Matematica e Informatica, Via G. Saragat 1, 44122 Ferrara, Italy }
\affiliation{INFN Sezione di Ferrara, Via G. Saragat 1, 44122 Ferrara, Italy}


\begin{abstract}

We report about the experimental set-up designed for the development of an innovative particle detector based on solid crystals of inert gases. The hybrid detection scheme, that exploits the electrons emission through the solid-vacuum interface, is demonstrated in matrices of solid neon and solid methane.

\end{abstract}

%Use showkeys class option if keyword
\maketitle

\section{Introduction}

Crystals made of inert gasses solidified at cryogenic temperature have been used since the 50's to study reactive species in a low interacting environment\,\cite{barnes2012, momose1998}. In this technique, named matrix isolation spectroscopy, the guest particles (atoms, molecules or ions) are embedded in a continuous matrix of solid crystal. As the matrix is made of un-reactive materials, diffusion processes are suppressed and only feeble interaction between host and guest can take place\,\cite{forstmann1980}. The dopant is then said to be isolated within the matrix and the guests atoms can be treated as free particles. This technique has been largely applied to study the spectra of a great number of stable and unstable atomic\,\cite{bohmer1980,silverman1997} and molecular\,\cite{jacox1989,jacox1994,weiss2016} species with the advantage of higher density with respect to spectroscopic studies in the gas phase. 
Recently this kind of materials have been also applied in different fields of physics both for fundamental studies and for applications. 
A search for the electric dipole moment of the electron was proposed exploiting a great number of strongly polar molecules embedded into solid crystals to reach high sensitivity in tests for symmetries violation\,\cite{pryor1987,kozlov2006,meyer2009}. The possibility to manipulate the interference of wave functions in a bulk solid was also demonstrated in a para--hydrogen matrix\,\cite{katsuki2013}.
Furthermore, applications as magnetic sensors, as qubits in quantum information and as platform for quantum simulation, have been investigated exploiting the long coherence time of electronic levels in high density spin materials implanted into the matrix\,\cite{upadhyay2016,kanagin2013,childress2014,kanagin2013,lemeshko2013,xu2011}. 

We believe that matrix isolation materials, combined with laser spectroscopy techniques, can be used to develop a new generation of low threshold particle detectors aimed to low interacting particle searches\,\cite{freedman1974} and dark matter studies\,\cite{dark_matter2016,dark_matter1_2012}.
The devised scheme is based on the possibility to extract electrons through the matrix-vacuum interface and to efficiently detect them in vacuum, owing to the high charge gain and low dark count rate of microchannel plate or semiconductor detectors\,\cite{fraser1987}. The emission process is instead related to the ground state energy of the electron within the matrix that is positive or compatible with zero, for instance, in solid neon (s-Ne)\,\cite{bolozdynya1999} and in methane (s-CH$_4$)\,\cite{asaf1983}. 
Beside the possibility to directly ionize the matrix, whose energy band gap is in the order of tens of electronvolt for s-Ne and s-CH$_4$, doping with alkali or rare earth would allow generation of charge carriers even for energy depositions much smaller that the matrix band gap. 
In fact with these guest atoms, that introduced sub--eV atomic energy levels, an intrinsic threshold $\Delta E$ can be engineered, provided the lacking ionization energy is given by a narrow bandwidth laser.
Furthermore,  $\Delta E$ coincides with the ground state Zeeman splitting in the presence of an external magnetic field, that can be tuned to match a precise energy range in axionic dark matter resonant searches \,\cite{sikivie2014,borghesani2015,santamaria2015}. For example, for a 10\,T magnetic field, axion masses in the order of $\sim$100\,$\mu$eV might be searched.
With undoped matrices WIMPs--related searches might instead be addressed in a mass range that differs from experiments based on noble liquid technology\,\cite{noble_liquid2013}.

In the present work we describe our apparatus that allows high purity crystals growth and verify electrons emission through the solid-vacuum interface in s-Ne and s-CH$_4$. We use electrons generated by photoelectric effect on gold to test the signal readout electronics and to investigate charge collection efficiency in a solid neon and methane matrices.

\section{experimental set-up}\label{sec:setup}

The experimental set-up consists of the gas purification system and a cryostat chamber equipped with electron detectors. We will describe each part in detail in the next subsections.

\subsection{Gas system}

\begin{figure}[h!]
\begin{center}
\includegraphics[width=3.45in]{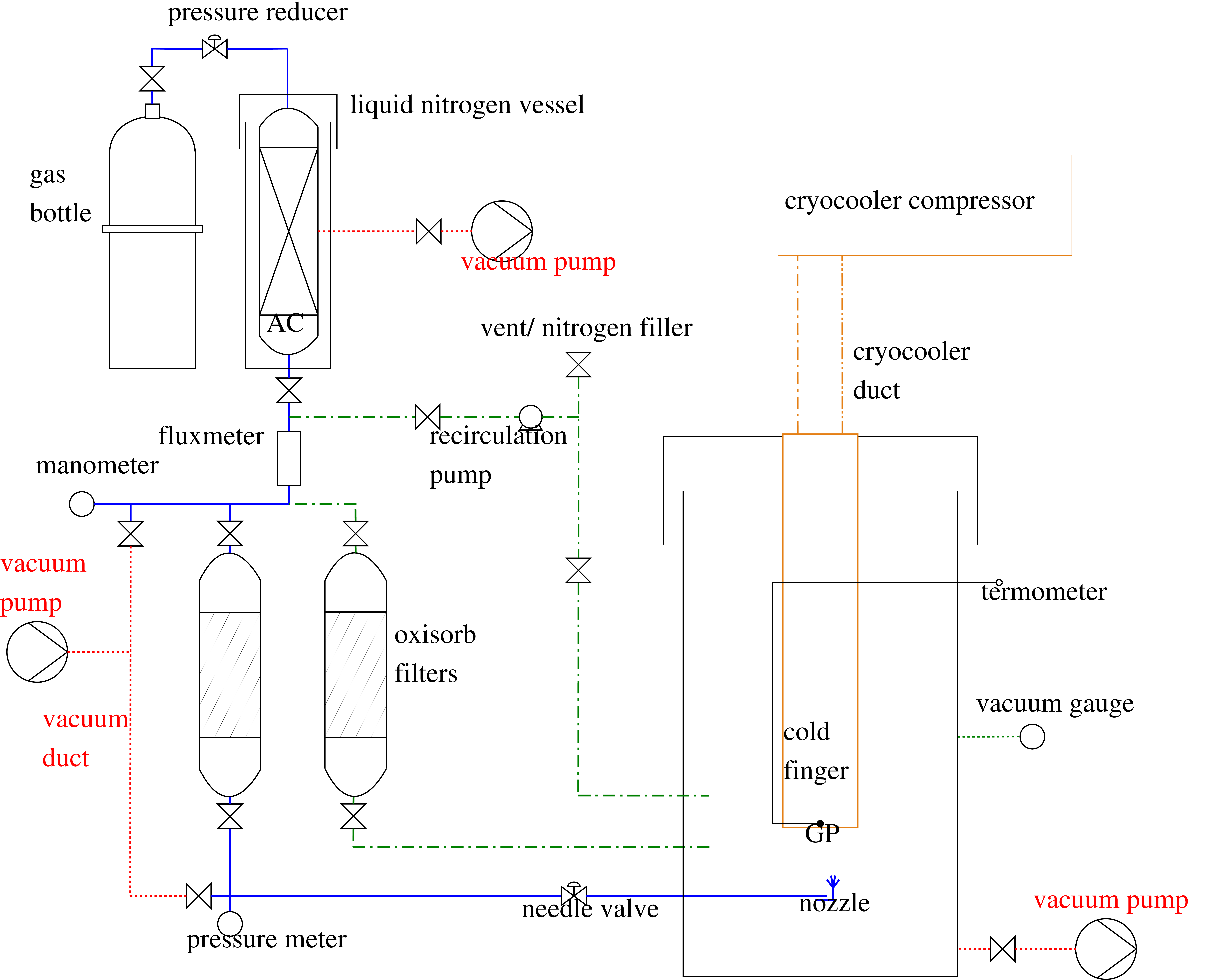}
\caption{(Colour online) Duct scheme of the set-up. Blue solid lines represent the CG line, green dashed lines are the PC line, red dotted lines stand for vacuum ducts.}
\label{fig:gas_system}
\end{center}
\end{figure}

Different gas bottles can be connected to the gas system shown in figure \ref{fig:gas_system}, depending on the material of the crystal that one wants to grow.  An activated charcoal filter (AC) is set after the gas bottle in which the starting material has a typical impurity concentration in the $\sim$ppm level. The AC trap is cooled down to cryogenic temperature (liquid nitrogen or other mixtures depending on the required temperature) in order to obtain a first step in the gas purification. Next to the AC, the pipe line is divided in two exclusive ways: one is used to purify the chamber (PC line) while the second is used for the crystal growth (CG line). 

The PC line is a loop line including the chamber and an Oxysorb\textsuperscript{\textregistered} large cartridge made of chromium embedded in a SiO$_2$ lattice that absorbs oxygen and moisture by a chemical process. Before the crystal growth the cryostat chamber walls are baked at 330\,K while the gas flows in the PC line for hours following the procedure described in reference\,\cite{carugno1996}.

The CG line starts after the AC filter, where a valve allows to isolate the PC line. An Oxysorb\textsuperscript{\textregistered} cartridge, identical to the one used in the PC line, is directly connected to the chamber through a needle valve which is necessary to finely control  the gas flux. The CG line ends inside the cryostat with a nozzle and a pressure meter is inserted to set the gas pressure during the crystal growth. 

We used all-metal bakeable Nupro\textsuperscript{\textregistered} valves with ConFlat\textsuperscript{\textregistered} flanges connected to adjacent components with copper O-rings to avoid possible contamination. 

With the described components a level of gas purification better than $5$\,ppb for O$_2$ impurities is expected\,\cite{carugno1996}.

\subsection{Cryostat}
The core of the set-up is a cryostat chamber where the crystals are grown and are maintained at low temperature. It consists of a four-hole cross stainless steel chamber equipped with four DN 200CF ConFlat\textsuperscript{\textregistered} flanges.  A two stages Helium pulse tube (Sumitomo RP82B2) allows cooling down to 4\,K temperature. The cold finger is enclosed in a copper shield maintained at 77\,K covered with five foils of mylar\textsuperscript{\textregistered} to reduce radiation heating. The front side of the chamber is equipped with a vacuum manipulator that allows to move both the gas nozzle and the charge detectors inside the chamber.
Future spectroscopic studies of doped crystals will be possible via a silica window mounted at the rear hole of the chamber.
The fourth access is connected to the vacuum system and a pressure lower than 10$^{-7}$\,mbar is maintained during the measurements.

\begin{figure}[h!]
\begin{center}
\includegraphics[width=3.45in]{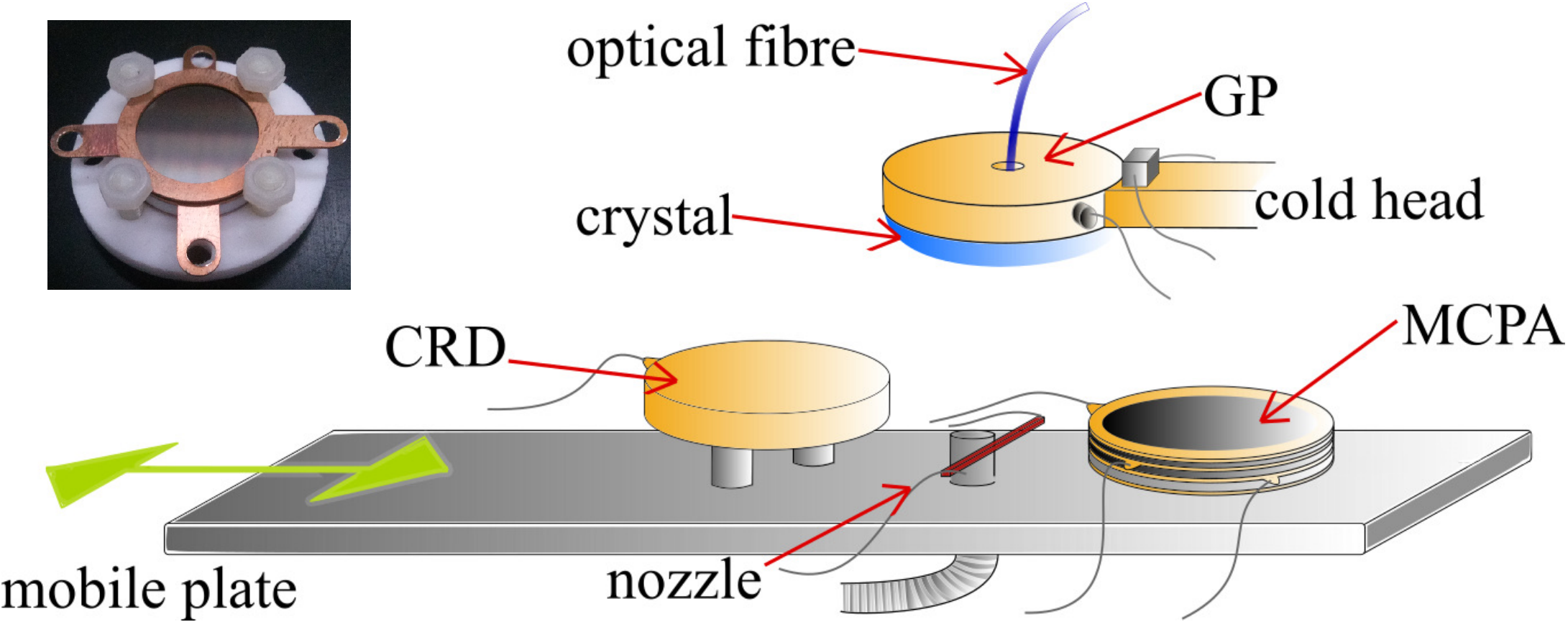}
\caption{(Colour online) Picture of the cold finger assembly and the mobile plate. The distance between the GP and the electrons detectors is $10\,$mm. In the inset is shown the micro channel plate assembly (MCPA).}
\label{fig:cryo}
\end{center}
\end{figure}

As shown in the figure \ref{fig:cryo}, the growth plate (GP) consists of a 25\,mm--diameter oxygen free, high thermal conductivity copper ring. The inner 8\,mm--diameter hole is covered with a gold foil of 100\,nm-thickness which is evaporated on a fused silica window. Electrons are generated by photoextraction that takes place when the frequency--quadrupled Nd:YAG ($\lambda=$266\,nm) output pulses impinge on the gold foil\,\cite{fischer1988}. As shown in figure\,\ref{fig:cryo} optical access to the cryostat is provided through a 3\,m--long fused silica fibre. 
The GP is fixed at the bottom of the cold finger and a good thermal contact between the different parts is ensured through an indium gasket. Fine control of the GP temperature is accomplished by Joule heating a 1\,Ohm electrical resistance and it is  monitored through a silicon diode mounted in the proximity of the crystal growth plate. 

Both the gas nozzle and the electron detectors are mounted on a mobile plate connected to a manipulator to place one or the other in front of the cathode growth plate. 
The gas nozzle is a 2\,mm--diameter hole at 30\,mm distance from the cold finger head. For the doped crystals growth we envisage to mount different alkali getter beside the jet.

We mounted two kinds of electrons detectors: a simple charge receiver disk (CRD) for tests involving more than$>10^5$ electrons, and a micro channel plate assembly (MCPA) when a much lower threshold detection is required.
The CRD plate consists of a 25\,mm--diameter copper disk, isolated from the ground and connected by a vacuum feedthrough to an external SHV port. The MCPA consists of one or double  $\pi$\,cm$^2$--area Hamamatsu F1094 micro-channel-plate interlaced with copper rings connected to an external HV power supply. The MCPA readout is ensured by a copper receiver disk mounted into a teflon cylinder.

\section{results}

Before the crystal growth, we test the field assisted gold photoextraction process in vacuum. As the quantum efficiency of the Au photoextraction process is $\sim10^{-5}$\,\cite{fischer1988}, we can estimate a maximum number of $\sim$10$^8$ generated electrons for a 1\,mJ laser pulse energy.

\begin{figure}[h!]
\begin{center}
\includegraphics[width=3.45in]{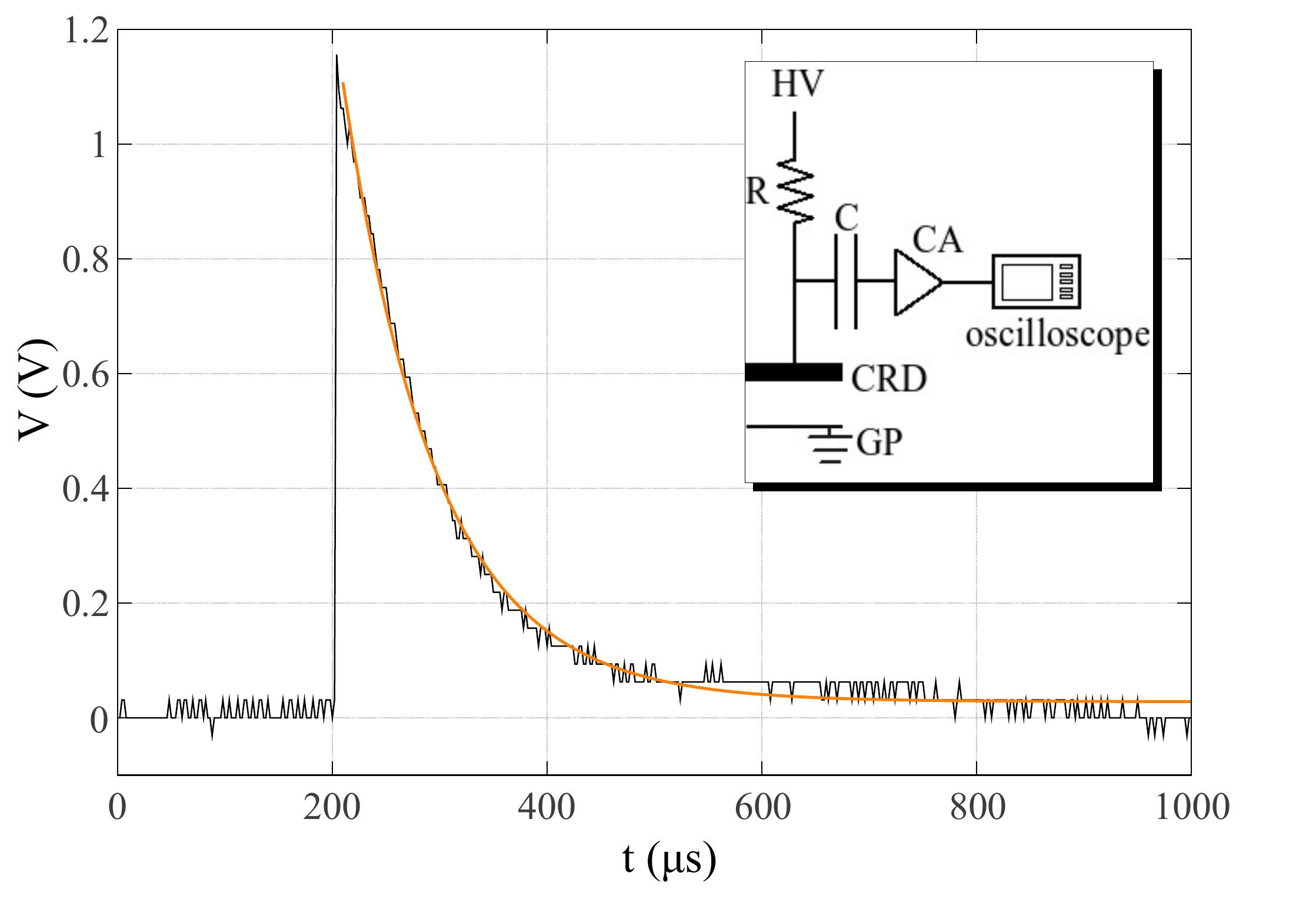}
\caption{(Colour online) Photo-current signal obtained at the oscilloscope. At t=200\,$\mu$s occurs the charge injection in the anode plate which is triggered by the laser pulse. Orange line is the fitting curve $y_0+A\cdot \exp({-t/\tau})$ with the free parameter $A$, $y_0$, $\tau$. The  inset is the electrical scheme to collect the signal.}
\label{fig:signal}
\end{center}
\end{figure}

As shown in the inset of figure\,\ref{fig:signal} the charge signal is picked up at a capacitor connected to the CRD and then amplified by a low noise charge amplifier (CA) with a gain G=1.25\,mV/fC. The charge signal registered at the oscilloscope is shown in figure  \ref{fig:signal}.
It can be observed that the signal increases rapidly in correspondence of the injection of charge and then it slowly decreases with the charge amplifier decay time $\sim90\,\mu$s. 

\begin{figure}[h!]
\begin{center}
\includegraphics[width=3.45in]{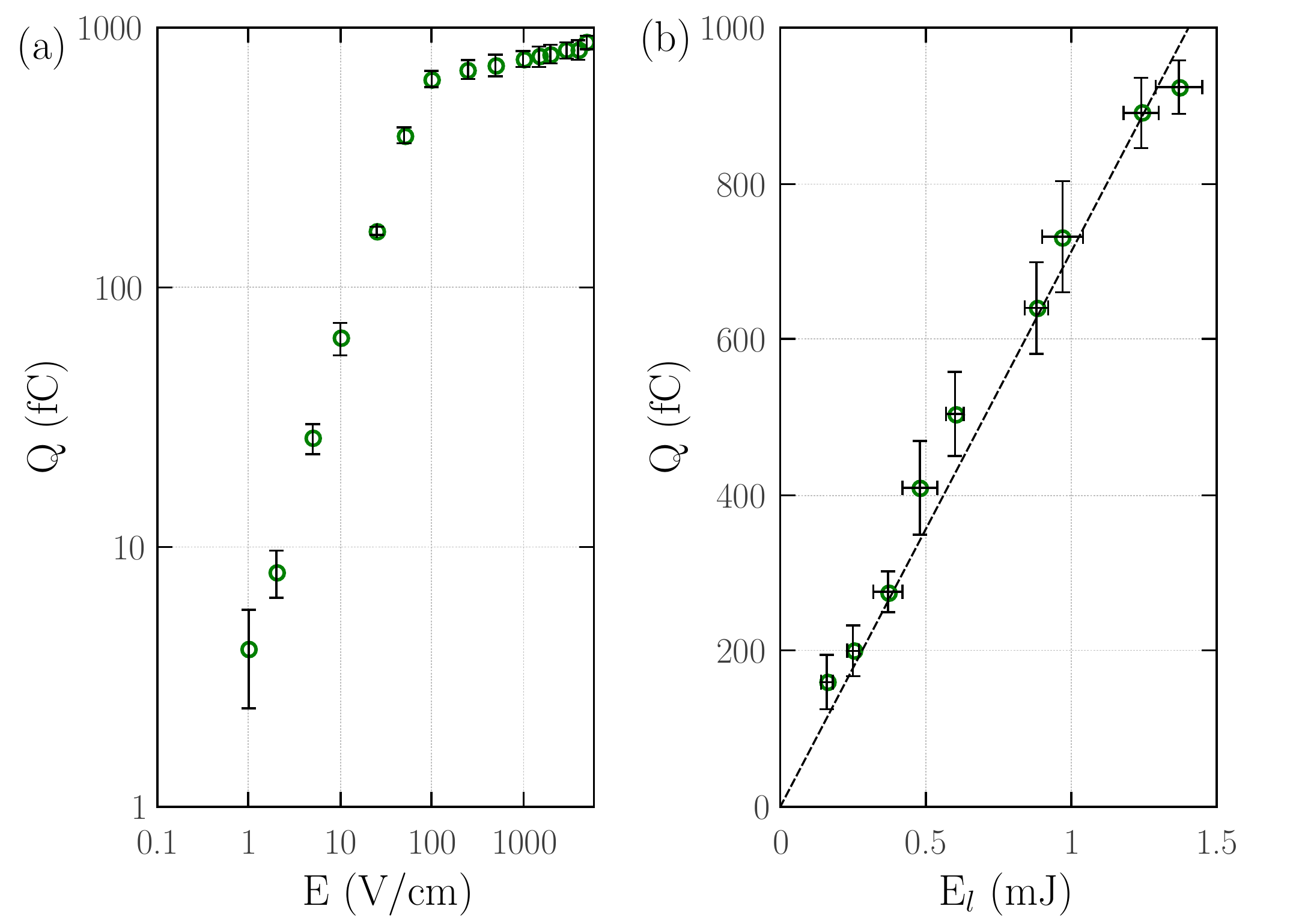}
\caption{a) Charge collected Q in vacuum for several values of the electric field strength E. b) Q plotted with respect to the energy of the laser pulses E$_l$ for E=500\,V/cm. }
\label{fig:tests}
\end{center}
\end{figure}

The collected charge data are plotted in figure\,\ref{fig:tests}(a) for several values of the electric field E at a fixed laser pulse energy E$_l$=1\,mJ. The measurement has been performed at room temperature with the chamber maintained at 2$\cdot 10^{-7}$\,mbar pressure. Full charge collection is accomplished for applied electric field greater than $100$\,V/cm.
Linearity of the collected charge Q with the laser pulse energy is demonstrated  in figure\,\ref{fig:tests}(b).

Crystal growth occurs through spraying the purified gas on the GP cathode maintained at a fixed temperature of 9\,K.
To ensure a high optical quality of the crystal the deposition rate is 0.2\,l/min and the pressure inside the chamber is kept at $\sim3\cdot$10$^{-5}$\,mbar. Growth is interrupted when the crystal layer is about 2\,mm and the temperature is then lowered at 4\,K after a few minutes of annealing.

\begin{figure}[h!]
\begin{center}
\includegraphics[width=3.45in]{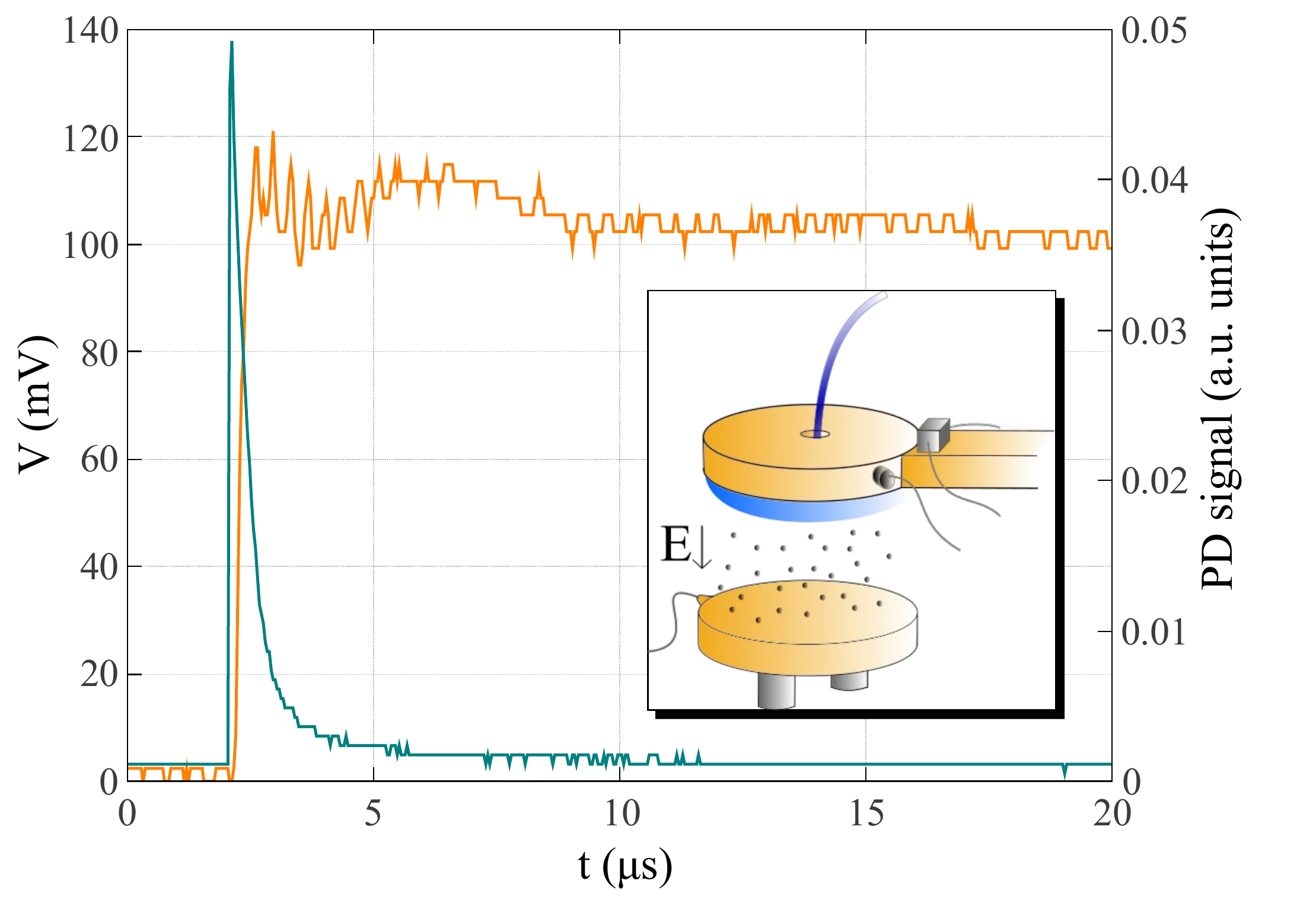}
\caption{(Colour online) Charge signal collected at CRD for electrons injected in a 2\,mm thickness Ne crystal (left scale). Also the optical trigger signal is shown (right scale). In the inset a picture of the crystal and the CRD. E=1\,kV/cm}
\label{fig:signalNe}
\end{center}
\end{figure}

In figure\,\ref{fig:signalNe} is shown the charge collection signal obtained in a s--Ne crystal when 1\,kV/cm is applied.
Differently from the previously described measurements the photoextraction takes place across the interface gold-crystal and electrons drift within the matrix with mobility in the order of 600\,cm$^2$V$^{-1}$s$^{-1}$\,\cite{sakai1982}.

We verified that the signal amplitude does not change through several hours under 1\,Hz--repetition rate photoinjections in the crystal.
Accordingly, no space charge effect takes place and this indicates that the overall charge collection process is efficient.
Similar signals were also obtained for electrons emission through the solid-vacuum  interface in methane crystals. 

\section{conclusions}\label{sec:concl}

In this work we have presented an innovative set-up developed for the growth and the study of inert gasses doped crystals. Such apparatus is devised for low energy threshold detection required in searches of low rate events. Potential applications of the system that we are developing, are the studies of dark matter axions,\cite{sikivie2014} and the research on neutrino physics besides all the application which requires detectors with low energy threshold and sensitive to low rate events\cite{bolozdynya1999}. Solid crystals doped at $\sim\%$ leads indeed to an improvement of 5 orders of magnitude in the target atom density as compared to a gas system\,\cite{santamaria2015}.

It has been demonstrated that the solid matrix can be grown with a high purity level, a necessary requirement in the described schemes for detection, i.e. the direct matrix ionization by the incident particle or the hybrid scheme that involves doped crystals probed by narrow linewidth laser.
Both the detection schemes are based on the extraction of the charges generated by the particle across the crystal--vacuum interface, a phenomenon that is assisted by an external electric field. High sensitivity electrons detectors are then used to collect single electrons in vacuum.
We have demonstrated efficiency of the charge collection process, by injecting into the crystals the electrons generated via photoelectric effect in a gold foil. Electric field--assisted electrons emission across solid--vacuum interface in s--Ne and s--CH$_4$ crystals of a few mm thickness has been observed in our measurements.

\section{acknowledgments}
This work is supported by the Istituto Nazionale di Fisica Nucleare under AXIOMA project. The authors wish to thank A. Bolozdynya, Dr. F. Chiossi, Prof. A.F. Borghesani, Prof. F. Pietropaolo, Prof. W. F. Schmidt and Prof B. Tremblay for the helpful discussions. For the technical support we thank Mr. L. Barcellan, Mr. E. Berto, Mr. F. Calaon, M. Rampazzo and M. Rebeschini. M. Guarise wish to thank Prof. T. Momose and his laboratory staff for precious suggestions.

%\nocite{*}
\bibliography{Matrix.bib}% Produces the bibliography via BibTeX.

\begin{thebibliography}{30}%
\makeatletter
\providecommand \@ifxundefined [1]{%
 \@ifx{#1\undefined}
}%
\providecommand \@ifnum [1]{%
 \ifnum #1\expandafter \@firstoftwo
 \else \expandafter \@secondoftwo
 \fi
}%
\providecommand \@ifx [1]{%
 \ifx #1\expandafter \@firstoftwo
 \else \expandafter \@secondoftwo
 \fi
}%
\providecommand \natexlab [1]{#1}%
\providecommand \enquote  [1]{``#1''}%
\providecommand \bibnamefont  [1]{#1}%
\providecommand \bibfnamefont [1]{#1}%
\providecommand \citenamefont [1]{#1}%
\providecommand \href@noop [0]{\@secondoftwo}%
\providecommand \href [0]{\begingroup \@sanitize@url \@href}%
\providecommand \@href[1]{\@@startlink{#1}\@@href}%
\providecommand \@@href[1]{\endgroup#1\@@endlink}%
\providecommand \@sanitize@url [0]{\catcode `\\12\catcode `\$12\catcode
  `\&12\catcode `\#12\catcode `\^12\catcode `\_12\catcode `\%12\relax}%
\providecommand \@@startlink[1]{}%
\providecommand \@@endlink[0]{}%
\providecommand \url  [0]{\begingroup\@sanitize@url \@url }%
\providecommand \@url [1]{\endgroup\@href {#1}{\urlprefix }}%
\providecommand \urlprefix  [0]{URL }%
\providecommand \Eprint [0]{\href }%
\providecommand \doibase [0]{http://dx.doi.org/}%
\providecommand \selectlanguage [0]{\@gobble}%
\providecommand \bibinfo  [0]{\@secondoftwo}%
\providecommand \bibfield  [0]{\@secondoftwo}%
\providecommand \translation [1]{[#1]}%
\providecommand \BibitemOpen [0]{}%
\providecommand \bibitemStop [0]{}%
\providecommand \bibitemNoStop [0]{.\EOS\space}%
\providecommand \EOS [0]{\spacefactor3000\relax}%
\providecommand \BibitemShut  [1]{\csname bibitem#1\endcsname}%
\let\auto@bib@innerbib\@empty
%</preamble>
\bibitem [{\citenamefont {Barnes}\ \emph {et~al.}(2012)\citenamefont {Barnes},
  \citenamefont {Orville-Thomas}, \citenamefont {Gaufr{\`e}s},\ and\
  \citenamefont {M{\"u}ller}}]{barnes2012}%
  \BibitemOpen
  \bibfield  {author} {\bibinfo {author} {\bibfnamefont {A.}~\bibnamefont
  {Barnes}}, \bibinfo {author} {\bibfnamefont {W.}~\bibnamefont
  {Orville-Thomas}}, \bibinfo {author} {\bibfnamefont {R.}~\bibnamefont
  {Gaufr{\`e}s}}, \ and\ \bibinfo {author} {\bibfnamefont {A.}~\bibnamefont
  {M{\"u}ller}},\ }\href@noop {} {\emph {\bibinfo {title} {{Matrix isolation
  spectroscopy}}}},\ Vol.~\bibinfo {volume} {76}\ (\bibinfo  {publisher}
  {Springer Science \& Business Media},\ \bibinfo {year} {2012})\BibitemShut
  {NoStop}%
\bibitem [{\citenamefont {Momose}\ and\ \citenamefont
  {Shida}(1998)}]{momose1998}%
  \BibitemOpen
  \bibfield  {author} {\bibinfo {author} {\bibfnamefont {T.}~\bibnamefont
  {Momose}}\ and\ \bibinfo {author} {\bibfnamefont {T.}~\bibnamefont {Shida}},\
  }\bibfield  {title} {\enquote {\bibinfo {title} {Matrix-isolation
  spectroscopy using solid parahydrogen as the matrix: application to
  high-resolution spectroscopy, photochemistry, and cryochemistry},}\
  }\href@noop {} {\bibfield  {journal} {\bibinfo  {journal} {Bulletin of the
  Chemical Society of Japan}\ }\textbf {\bibinfo {volume} {71}},\ \bibinfo
  {pages} {1--15} (\bibinfo {year} {1998})}\BibitemShut {NoStop}%
\bibitem [{\citenamefont {Forstmann}\ and\ \citenamefont
  {Ossicini}(1980)}]{forstmann1980}%
  \BibitemOpen
  \bibfield  {author} {\bibinfo {author} {\bibfnamefont {F.}~\bibnamefont
  {Forstmann}}\ and\ \bibinfo {author} {\bibfnamefont {S.}~\bibnamefont
  {Ossicini}},\ }\bibfield  {title} {\enquote {\bibinfo {title} {{The influence
  of a rare-gas matrix on the electronic levels of isolated atoms}},}\
  }\href@noop {} {\bibfield  {journal} {\bibinfo  {journal} {The Journal of
  Chemical Physics}\ }\textbf {\bibinfo {volume} {73}},\ \bibinfo {pages}
  {5997--6002} (\bibinfo {year} {1980})}\BibitemShut {NoStop}%
\bibitem [{\citenamefont {B{\"o}hmer}\ \emph {et~al.}(1980)\citenamefont
  {B{\"o}hmer}, \citenamefont {Haensel}, \citenamefont {Schwentner},\ and\
  \citenamefont {Boursey}}]{bohmer1980}%
  \BibitemOpen
  \bibfield  {author} {\bibinfo {author} {\bibfnamefont {W.}~\bibnamefont
  {B{\"o}hmer}}, \bibinfo {author} {\bibfnamefont {R.}~\bibnamefont {Haensel}},
  \bibinfo {author} {\bibfnamefont {N.}~\bibnamefont {Schwentner}}, \ and\
  \bibinfo {author} {\bibfnamefont {E.}~\bibnamefont {Boursey}},\ }\bibfield
  {title} {\enquote {\bibinfo {title} {Excitation and emission bands of
  hydrogen atoms in a solid neon matrix},}\ }\href@noop {} {\bibfield
  {journal} {\bibinfo  {journal} {Chemical Physics}\ }\textbf {\bibinfo
  {volume} {49}},\ \bibinfo {pages} {225--230} (\bibinfo {year}
  {1980})}\BibitemShut {NoStop}%
\bibitem [{\citenamefont {Silverman}\ and\ \citenamefont
  {Fajardo}(1997)}]{silverman1997}%
  \BibitemOpen
  \bibfield  {author} {\bibinfo {author} {\bibfnamefont {D.~C.}\ \bibnamefont
  {Silverman}}\ and\ \bibinfo {author} {\bibfnamefont {M.~E.}\ \bibnamefont
  {Fajardo}},\ }\bibfield  {title} {\enquote {\bibinfo {title} {Matrix
  isolation spectroscopy of na atoms deposited as na+ ions},}\ }\href@noop {}
  {\bibfield  {journal} {\bibinfo  {journal} {The Journal of chemical physics}\
  }\textbf {\bibinfo {volume} {106}},\ \bibinfo {pages} {8964--8966} (\bibinfo
  {year} {1997})}\BibitemShut {NoStop}%
\bibitem [{\citenamefont {Jacox}\ and\ \citenamefont
  {Thompson}(1989)}]{jacox1989}%
  \BibitemOpen
  \bibfield  {author} {\bibinfo {author} {\bibfnamefont {M.~E.}\ \bibnamefont
  {Jacox}}\ and\ \bibinfo {author} {\bibfnamefont {W.~E.}\ \bibnamefont
  {Thompson}},\ }\bibfield  {title} {\enquote {\bibinfo {title} {The
  vibrational spectra of molecular ions isolated in solid neon. i. co+ 2 and
  co- 2},}\ }\href@noop {} {\bibfield  {journal} {\bibinfo  {journal} {The
  Journal of chemical physics}\ }\textbf {\bibinfo {volume} {91}},\ \bibinfo
  {pages} {1410--1416} (\bibinfo {year} {1989})}\BibitemShut {NoStop}%
\bibitem [{\citenamefont {Jacox}(1994)}]{jacox1994}%
  \BibitemOpen
  \bibfield  {author} {\bibinfo {author} {\bibfnamefont {M.~E.}\ \bibnamefont
  {Jacox}},\ }\bibfield  {title} {\enquote {\bibinfo {title} {The vibrational
  energy levels of small transient molecules isolated in neon and argon
  matrices},}\ }\href@noop {} {\bibfield  {journal} {\bibinfo  {journal}
  {Chemical physics}\ }\textbf {\bibinfo {volume} {189}},\ \bibinfo {pages}
  {149--170} (\bibinfo {year} {1994})}\BibitemShut {NoStop}%
\bibitem [{\citenamefont {Weiss}, \citenamefont {Waller},\ and\ \citenamefont
  {Phillips}(2016)}]{weiss2016}%
  \BibitemOpen
  \bibfield  {author} {\bibinfo {author} {\bibfnamefont {N.~M.}\ \bibnamefont
  {Weiss}}, \bibinfo {author} {\bibfnamefont {A.~W.}\ \bibnamefont {Waller}}, \
  and\ \bibinfo {author} {\bibfnamefont {J.~A.}\ \bibnamefont {Phillips}},\
  }\bibfield  {title} {\enquote {\bibinfo {title} {Infrared spectrum of ch 3
  cn--hcl in solid neon, and modeling matrix effects in ch 3 cn--hcl and h 3
  n--hcl},}\ }\href@noop {} {\bibfield  {journal} {\bibinfo  {journal} {Journal
  of Molecular Structure}\ }\textbf {\bibinfo {volume} {1105}},\ \bibinfo
  {pages} {341--349} (\bibinfo {year} {2016})}\BibitemShut {NoStop}%
\bibitem [{\citenamefont {Pryor}\ and\ \citenamefont
  {Wilczek}(1987)}]{pryor1987}%
  \BibitemOpen
  \bibfield  {author} {\bibinfo {author} {\bibfnamefont {C.}~\bibnamefont
  {Pryor}}\ and\ \bibinfo {author} {\bibfnamefont {F.}~\bibnamefont
  {Wilczek}},\ }\bibfield  {title} {\enquote {\bibinfo {title} {“artificial
  vacuum” for t-violation experiment},}\ }\href@noop {} {\bibfield  {journal}
  {\bibinfo  {journal} {Physics Letters B}\ }\textbf {\bibinfo {volume}
  {194}},\ \bibinfo {pages} {137--140} (\bibinfo {year} {1987})}\BibitemShut
  {NoStop}%
\bibitem [{\citenamefont {Kozlov}\ and\ \citenamefont
  {Derevianko}(2006)}]{kozlov2006}%
  \BibitemOpen
  \bibfield  {author} {\bibinfo {author} {\bibfnamefont {M.}~\bibnamefont
  {Kozlov}}\ and\ \bibinfo {author} {\bibfnamefont {A.}~\bibnamefont
  {Derevianko}},\ }\bibfield  {title} {\enquote {\bibinfo {title} {Proposal for
  a sensitive search for the electric dipole moment of the electron with
  matrix-isolated radicals},}\ }\href@noop {} {\bibfield  {journal} {\bibinfo
  {journal} {Physical review letters}\ }\textbf {\bibinfo {volume} {97}},\
  \bibinfo {pages} {063001} (\bibinfo {year} {2006})}\BibitemShut {NoStop}%
\bibitem [{\citenamefont {Meyer}\ and\ \citenamefont {Bohn}(2009)}]{meyer2009}%
  \BibitemOpen
  \bibfield  {author} {\bibinfo {author} {\bibfnamefont {E.~R.}\ \bibnamefont
  {Meyer}}\ and\ \bibinfo {author} {\bibfnamefont {J.~L.}\ \bibnamefont
  {Bohn}},\ }\bibfield  {title} {\enquote {\bibinfo {title} {Electron
  electric-dipole-moment searches based on alkali-metal-or
  alkaline-earth-metal-bearing molecules},}\ }\href@noop {} {\bibfield
  {journal} {\bibinfo  {journal} {Physical Review A}\ }\textbf {\bibinfo
  {volume} {80}},\ \bibinfo {pages} {042508} (\bibinfo {year}
  {2009})}\BibitemShut {NoStop}%
\bibitem [{\citenamefont {Katsuki}, \citenamefont {Kayanuma},\ and\
  \citenamefont {Ohmori}(2013)}]{katsuki2013}%
  \BibitemOpen
  \bibfield  {author} {\bibinfo {author} {\bibfnamefont {H.}~\bibnamefont
  {Katsuki}}, \bibinfo {author} {\bibfnamefont {Y.}~\bibnamefont {Kayanuma}}, \
  and\ \bibinfo {author} {\bibfnamefont {K.}~\bibnamefont {Ohmori}},\
  }\bibfield  {title} {\enquote {\bibinfo {title} {Optically engineered quantum
  interference of delocalized wave functions in a bulk solid: The example of
  solid para-hydrogen},}\ }\href@noop {} {\bibfield  {journal} {\bibinfo
  {journal} {Physical Review B}\ }\textbf {\bibinfo {volume} {88}},\ \bibinfo
  {pages} {014507} (\bibinfo {year} {2013})}\BibitemShut {NoStop}%
\bibitem [{\citenamefont {Upadhyay}\ \emph {et~al.}(2016)\citenamefont
  {Upadhyay}, \citenamefont {Kanagin}, \citenamefont {Hartzell}, \citenamefont
  {Christy}, \citenamefont {Arnott}, \citenamefont {Momose}, \citenamefont
  {Patterson},\ and\ \citenamefont {Weinstein}}]{upadhyay2016}%
  \BibitemOpen
  \bibfield  {author} {\bibinfo {author} {\bibfnamefont {S.}~\bibnamefont
  {Upadhyay}}, \bibinfo {author} {\bibfnamefont {A.~N.}\ \bibnamefont
  {Kanagin}}, \bibinfo {author} {\bibfnamefont {C.}~\bibnamefont {Hartzell}},
  \bibinfo {author} {\bibfnamefont {T.}~\bibnamefont {Christy}}, \bibinfo
  {author} {\bibfnamefont {W.~P.}\ \bibnamefont {Arnott}}, \bibinfo {author}
  {\bibfnamefont {T.}~\bibnamefont {Momose}}, \bibinfo {author} {\bibfnamefont
  {D.}~\bibnamefont {Patterson}}, \ and\ \bibinfo {author} {\bibfnamefont
  {J.~D.}\ \bibnamefont {Weinstein}},\ }\bibfield  {title} {\enquote {\bibinfo
  {title} {Longitudinal spin relaxation of optically pumped rubidium atoms in
  solid parahydrogen},}\ }\href@noop {} {\bibfield  {journal} {\bibinfo
  {journal} {Physical review letters}\ }\textbf {\bibinfo {volume} {117}},\
  \bibinfo {pages} {175301} (\bibinfo {year} {2016})}\BibitemShut {NoStop}%
\bibitem [{\citenamefont {Kanagin}\ \emph {et~al.}(2013)\citenamefont
  {Kanagin}, \citenamefont {Regmi}, \citenamefont {Pathak},\ and\ \citenamefont
  {Weinstein}}]{kanagin2013}%
  \BibitemOpen
  \bibfield  {author} {\bibinfo {author} {\bibfnamefont {A.~N.}\ \bibnamefont
  {Kanagin}}, \bibinfo {author} {\bibfnamefont {S.~K.}\ \bibnamefont {Regmi}},
  \bibinfo {author} {\bibfnamefont {P.}~\bibnamefont {Pathak}}, \ and\ \bibinfo
  {author} {\bibfnamefont {J.~D.}\ \bibnamefont {Weinstein}},\ }\bibfield
  {title} {\enquote {\bibinfo {title} {Optical pumping of rubidium atoms frozen
  in solid argon},}\ }\href@noop {} {\bibfield  {journal} {\bibinfo  {journal}
  {Physical Review A}\ }\textbf {\bibinfo {volume} {88}},\ \bibinfo {pages}
  {063404} (\bibinfo {year} {2013})}\BibitemShut {NoStop}%
\bibitem [{\citenamefont {Childress}, \citenamefont {Walsworth},\ and\
  \citenamefont {Lukin}(2014)}]{childress2014}%
  \BibitemOpen
  \bibfield  {author} {\bibinfo {author} {\bibfnamefont {L.}~\bibnamefont
  {Childress}}, \bibinfo {author} {\bibfnamefont {R.}~\bibnamefont
  {Walsworth}}, \ and\ \bibinfo {author} {\bibfnamefont {M.}~\bibnamefont
  {Lukin}},\ }\bibfield  {title} {\enquote {\bibinfo {title} {Atom-like crystal
  defects: From quantum computers to biological sensors},}\ }\href@noop {}
  {\bibfield  {journal} {\bibinfo  {journal} {Physics Today}\ }\textbf
  {\bibinfo {volume} {67}},\ \bibinfo {pages} {38--43} (\bibinfo {year}
  {2014})}\BibitemShut {NoStop}%
\bibitem [{\citenamefont {Lemeshko}\ \emph {et~al.}(2013)\citenamefont
  {Lemeshko}, \citenamefont {Yao}, \citenamefont {Gorshkov}, \citenamefont
  {Weimer}, \citenamefont {Bennett}, \citenamefont {Momose},\ and\
  \citenamefont {Gopalakrishnan}}]{lemeshko2013}%
  \BibitemOpen
  \bibfield  {author} {\bibinfo {author} {\bibfnamefont {M.}~\bibnamefont
  {Lemeshko}}, \bibinfo {author} {\bibfnamefont {N.~Y.}\ \bibnamefont {Yao}},
  \bibinfo {author} {\bibfnamefont {A.~V.}\ \bibnamefont {Gorshkov}}, \bibinfo
  {author} {\bibfnamefont {H.}~\bibnamefont {Weimer}}, \bibinfo {author}
  {\bibfnamefont {S.~D.}\ \bibnamefont {Bennett}}, \bibinfo {author}
  {\bibfnamefont {T.}~\bibnamefont {Momose}}, \ and\ \bibinfo {author}
  {\bibfnamefont {S.}~\bibnamefont {Gopalakrishnan}},\ }\bibfield  {title}
  {\enquote {\bibinfo {title} {Controllable quantum spin glasses with magnetic
  impurities embedded in quantum solids},}\ }\href@noop {} {\bibfield
  {journal} {\bibinfo  {journal} {Physical Review B}\ }\textbf {\bibinfo
  {volume} {88}},\ \bibinfo {pages} {014426} (\bibinfo {year}
  {2013})}\BibitemShut {NoStop}%
\bibitem [{\citenamefont {Xu}\ \emph {et~al.}(2011)\citenamefont {Xu},
  \citenamefont {Hu}, \citenamefont {Singh}, \citenamefont {Bailey},
  \citenamefont {Lu}, \citenamefont {Mueller}, \citenamefont {O’Connor},
  \citenamefont {Welp} \emph {et~al.}}]{xu2011}%
  \BibitemOpen
  \bibfield  {author} {\bibinfo {author} {\bibfnamefont {C.-Y.}\ \bibnamefont
  {Xu}}, \bibinfo {author} {\bibfnamefont {S.-M.}\ \bibnamefont {Hu}}, \bibinfo
  {author} {\bibfnamefont {J.}~\bibnamefont {Singh}}, \bibinfo {author}
  {\bibfnamefont {K.}~\bibnamefont {Bailey}}, \bibinfo {author} {\bibfnamefont
  {Z.-T.}\ \bibnamefont {Lu}}, \bibinfo {author} {\bibfnamefont
  {P.}~\bibnamefont {Mueller}}, \bibinfo {author} {\bibfnamefont
  {T.}~\bibnamefont {O’Connor}}, \bibinfo {author} {\bibfnamefont
  {U.}~\bibnamefont {Welp}},  \emph {et~al.},\ }\bibfield  {title} {\enquote
  {\bibinfo {title} {Optical excitation and decay dynamics of ytterbium atoms
  embedded in a solid neon matrix},}\ }\href@noop {} {\bibfield  {journal}
  {\bibinfo  {journal} {Physical review letters}\ }\textbf {\bibinfo {volume}
  {107}},\ \bibinfo {pages} {093001} (\bibinfo {year} {2011})}\BibitemShut
  {NoStop}%
\bibitem [{\citenamefont {Freedman}(1974)}]{freedman1974}%
  \BibitemOpen
  \bibfield  {author} {\bibinfo {author} {\bibfnamefont {D.~Z.}\ \bibnamefont
  {Freedman}},\ }\bibfield  {title} {\enquote {\bibinfo {title} {Coherent
  effects of a weak neutral current},}\ }\href@noop {} {\bibfield  {journal}
  {\bibinfo  {journal} {Physical Review D}\ }\textbf {\bibinfo {volume} {9}},\
  \bibinfo {pages} {1389} (\bibinfo {year} {1974})}\BibitemShut {NoStop}%
\bibitem [{\citenamefont {Mayet}\ \emph {et~al.}(2016)\citenamefont {Mayet},
  \citenamefont {Green}, \citenamefont {Battat}, \citenamefont {Billard},
  \citenamefont {Bozorgnia}, \citenamefont {Gelmini}, \citenamefont {Gondolo},
  \citenamefont {Kavanagh}, \citenamefont {Lee}, \citenamefont {Loomba} \emph
  {et~al.}}]{dark_matter2016}%
  \BibitemOpen
  \bibfield  {author} {\bibinfo {author} {\bibfnamefont {F.}~\bibnamefont
  {Mayet}}, \bibinfo {author} {\bibfnamefont {A.~M.}\ \bibnamefont {Green}},
  \bibinfo {author} {\bibfnamefont {J.}~\bibnamefont {Battat}}, \bibinfo
  {author} {\bibfnamefont {J.}~\bibnamefont {Billard}}, \bibinfo {author}
  {\bibfnamefont {N.}~\bibnamefont {Bozorgnia}}, \bibinfo {author}
  {\bibfnamefont {G.}~\bibnamefont {Gelmini}}, \bibinfo {author} {\bibfnamefont
  {P.}~\bibnamefont {Gondolo}}, \bibinfo {author} {\bibfnamefont
  {B.}~\bibnamefont {Kavanagh}}, \bibinfo {author} {\bibfnamefont
  {S.}~\bibnamefont {Lee}}, \bibinfo {author} {\bibfnamefont {D.}~\bibnamefont
  {Loomba}},  \emph {et~al.},\ }\bibfield  {title} {\enquote {\bibinfo {title}
  {A review of the discovery reach of directional dark matter detection},}\
  }\href@noop {} {\bibfield  {journal} {\bibinfo  {journal} {Physics Reports}\
  }\textbf {\bibinfo {volume} {627}},\ \bibinfo {pages} {1--49} (\bibinfo
  {year} {2016})}\BibitemShut {NoStop}%
\bibitem [{\citenamefont {Baudis}(2012)}]{dark_matter1_2012}%
  \BibitemOpen
  \bibfield  {author} {\bibinfo {author} {\bibfnamefont {L.}~\bibnamefont
  {Baudis}},\ }\bibfield  {title} {\enquote {\bibinfo {title} {Direct dark
  matter detection: the next decade},}\ }\href@noop {} {\bibfield  {journal}
  {\bibinfo  {journal} {Physics of the Dark Universe}\ }\textbf {\bibinfo
  {volume} {1}},\ \bibinfo {pages} {94--108} (\bibinfo {year}
  {2012})}\BibitemShut {NoStop}%
\bibitem [{\citenamefont {Fraser}, \citenamefont {Pearson},\ and\ \citenamefont
  {Lees}(1987)}]{fraser1987}%
  \BibitemOpen
  \bibfield  {author} {\bibinfo {author} {\bibfnamefont {G.}~\bibnamefont
  {Fraser}}, \bibinfo {author} {\bibfnamefont {J.}~\bibnamefont {Pearson}}, \
  and\ \bibinfo {author} {\bibfnamefont {J.}~\bibnamefont {Lees}},\ }\bibfield
  {title} {\enquote {\bibinfo {title} {Dark noise in microchannel plate x-ray
  detectors},}\ }\href@noop {} {\bibfield  {journal} {\bibinfo  {journal}
  {Nuclear Instruments and Methods in Physics Research Section A: Accelerators,
  Spectrometers, Detectors and Associated Equipment}\ }\textbf {\bibinfo
  {volume} {254}},\ \bibinfo {pages} {447--462} (\bibinfo {year}
  {1987})}\BibitemShut {NoStop}%
\bibitem [{\citenamefont {Bolozdynya}(1999)}]{bolozdynya1999}%
  \BibitemOpen
  \bibfield  {author} {\bibinfo {author} {\bibfnamefont {A.}~\bibnamefont
  {Bolozdynya}},\ }\bibfield  {title} {\enquote {\bibinfo {title} {Two-phase
  emission detectors and their applications},}\ }\href@noop {} {\bibfield
  {journal} {\bibinfo  {journal} {Nuclear Instruments and Methods in Physics
  Research Section A: Accelerators, Spectrometers, Detectors and Associated
  Equipment}\ }\textbf {\bibinfo {volume} {422}},\ \bibinfo {pages} {314--320}
  (\bibinfo {year} {1999})}\BibitemShut {NoStop}%
\bibitem [{\citenamefont {Asaf}, \citenamefont {Reininger},\ and\ \citenamefont
  {Steinberger}(1983)}]{asaf1983}%
  \BibitemOpen
  \bibfield  {author} {\bibinfo {author} {\bibfnamefont {U.}~\bibnamefont
  {Asaf}}, \bibinfo {author} {\bibfnamefont {R.}~\bibnamefont {Reininger}}, \
  and\ \bibinfo {author} {\bibfnamefont {I.}~\bibnamefont {Steinberger}},\
  }\bibfield  {title} {\enquote {\bibinfo {title} {The energy v0 of the
  quasi-free electron in gaseous, liquid and solid methane},}\ }\href@noop {}
  {\bibfield  {journal} {\bibinfo  {journal} {Chemical physics letters}\
  }\textbf {\bibinfo {volume} {100}},\ \bibinfo {pages} {363--366} (\bibinfo
  {year} {1983})}\BibitemShut {NoStop}%
\bibitem [{\citenamefont {Sikivie}(2014)}]{sikivie2014}%
  \BibitemOpen
  \bibfield  {author} {\bibinfo {author} {\bibfnamefont {P.}~\bibnamefont
  {Sikivie}},\ }\bibfield  {title} {\enquote {\bibinfo {title} {Axion dark
  matter detection using atomic transitions},}\ }\href@noop {} {\bibfield
  {journal} {\bibinfo  {journal} {Physical review letters}\ }\textbf {\bibinfo
  {volume} {113}},\ \bibinfo {pages} {201301} (\bibinfo {year}
  {2014})}\BibitemShut {NoStop}%
\bibitem [{\citenamefont {Borghesani}\ \emph {et~al.}(2015)\citenamefont
  {Borghesani}, \citenamefont {Braggio}, \citenamefont {Carugno}, \citenamefont
  {Chiossi}, \citenamefont {{Di Lieto}}, \citenamefont {Guarise}, \citenamefont
  {Ruoso},\ and\ \citenamefont {Tonelli}}]{borghesani2015}%
  \BibitemOpen
  \bibfield  {author} {\bibinfo {author} {\bibfnamefont {A.}~\bibnamefont
  {Borghesani}}, \bibinfo {author} {\bibfnamefont {C.}~\bibnamefont {Braggio}},
  \bibinfo {author} {\bibfnamefont {G.}~\bibnamefont {Carugno}}, \bibinfo
  {author} {\bibfnamefont {F.}~\bibnamefont {Chiossi}}, \bibinfo {author}
  {\bibfnamefont {A.}~\bibnamefont {{Di Lieto}}}, \bibinfo {author}
  {\bibfnamefont {M.}~\bibnamefont {Guarise}}, \bibinfo {author} {\bibfnamefont
  {G.}~\bibnamefont {Ruoso}}, \ and\ \bibinfo {author} {\bibfnamefont
  {M.}~\bibnamefont {Tonelli}},\ }\bibfield  {title} {\enquote {\bibinfo
  {title} {{Particle detection through the quantum counter concept in YAG:
  Er3+}},}\ }\href@noop {} {\bibfield  {journal} {\bibinfo  {journal} {Applied
  Physics Letters}\ }\textbf {\bibinfo {volume} {107}},\ \bibinfo {pages}
  {193501} (\bibinfo {year} {2015})}\BibitemShut {NoStop}%
\bibitem [{\citenamefont {Santamaria}\ \emph {et~al.}(2015)\citenamefont
  {Santamaria}, \citenamefont {Braggio}, \citenamefont {Carugno}, \citenamefont
  {Di~Sarno}, \citenamefont {Maddaloni},\ and\ \citenamefont
  {Ruoso}}]{santamaria2015}%
  \BibitemOpen
  \bibfield  {author} {\bibinfo {author} {\bibfnamefont {L.}~\bibnamefont
  {Santamaria}}, \bibinfo {author} {\bibfnamefont {C.}~\bibnamefont {Braggio}},
  \bibinfo {author} {\bibfnamefont {G.}~\bibnamefont {Carugno}}, \bibinfo
  {author} {\bibfnamefont {V.}~\bibnamefont {Di~Sarno}}, \bibinfo {author}
  {\bibfnamefont {P.}~\bibnamefont {Maddaloni}}, \ and\ \bibinfo {author}
  {\bibfnamefont {G.}~\bibnamefont {Ruoso}},\ }\bibfield  {title} {\enquote
  {\bibinfo {title} {Axion dark matter detection by laser spectroscopy of
  ultracold molecular oxygen: a proposal},}\ }\href@noop {} {\bibfield
  {journal} {\bibinfo  {journal} {New Journal of Physics}\ }\textbf {\bibinfo
  {volume} {17}},\ \bibinfo {pages} {113025} (\bibinfo {year}
  {2015})}\BibitemShut {NoStop}%
\bibitem [{\citenamefont {Chepel}\ and\ \citenamefont
  {Ara{\'u}jo}(2013)}]{noble_liquid2013}%
  \BibitemOpen
  \bibfield  {author} {\bibinfo {author} {\bibfnamefont {V.}~\bibnamefont
  {Chepel}}\ and\ \bibinfo {author} {\bibfnamefont {H.}~\bibnamefont
  {Ara{\'u}jo}},\ }\bibfield  {title} {\enquote {\bibinfo {title} {Liquid noble
  gas detectors for low energy particle physics},}\ }\href@noop {} {\bibfield
  {journal} {\bibinfo  {journal} {Journal of Instrumentation}\ }\textbf
  {\bibinfo {volume} {8}},\ \bibinfo {pages} {R04001} (\bibinfo {year}
  {2013})}\BibitemShut {NoStop}%
\bibitem [{\citenamefont {Carugno}\ \emph {et~al.}(1996)\citenamefont
  {Carugno}, \citenamefont {Bressi}, \citenamefont {Cerdonio}, \citenamefont
  {Conti}, \citenamefont {Meneguzzo}, \citenamefont {Onofrio}, \citenamefont
  {Zanello}, \citenamefont {Beriotto}, \citenamefont {De~Biasia}, \citenamefont
  {Nicoletto} \emph {et~al.}}]{carugno1996}%
  \BibitemOpen
  \bibfield  {author} {\bibinfo {author} {\bibfnamefont {G.}~\bibnamefont
  {Carugno}}, \bibinfo {author} {\bibfnamefont {G.}~\bibnamefont {Bressi}},
  \bibinfo {author} {\bibfnamefont {S.}~\bibnamefont {Cerdonio}}, \bibinfo
  {author} {\bibfnamefont {E.}~\bibnamefont {Conti}}, \bibinfo {author}
  {\bibfnamefont {A.}~\bibnamefont {Meneguzzo}}, \bibinfo {author}
  {\bibfnamefont {R.}~\bibnamefont {Onofrio}}, \bibinfo {author} {\bibfnamefont
  {D.}~\bibnamefont {Zanello}}, \bibinfo {author} {\bibfnamefont
  {U.}~\bibnamefont {Beriotto}}, \bibinfo {author} {\bibfnamefont
  {S.}~\bibnamefont {De~Biasia}}, \bibinfo {author} {\bibfnamefont
  {M.}~\bibnamefont {Nicoletto}},  \emph {et~al.},\ }\bibfield  {title}
  {\enquote {\bibinfo {title} {A large liquid xenon time projection chamber for
  the study of the radiative pion decay},}\ }\href@noop {} {\bibfield
  {journal} {\bibinfo  {journal} {Nuclear Instruments and Methods in Physics
  Research Section A: Accelerators, Spectrometers, Detectors and Associated
  Equipment}\ }\textbf {\bibinfo {volume} {376}},\ \bibinfo {pages} {149--154}
  (\bibinfo {year} {1996})}\BibitemShut {NoStop}%
\bibitem [{\citenamefont {Fischer}\ and\ \citenamefont
  {Rao-Srinivasan}(1988)}]{fischer1988}%
  \BibitemOpen
  \bibfield  {author} {\bibinfo {author} {\bibfnamefont {J.}~\bibnamefont
  {Fischer}}\ and\ \bibinfo {author} {\bibfnamefont {T.}~\bibnamefont
  {Rao-Srinivasan}},\ }\href@noop {} {\enquote {\bibinfo {title} {Uv
  photoemission studies of metal photocathodes for particle accelerators},}\
  }\bibinfo {type} {Tech. Rep.}\ (\bibinfo  {institution} {Brookhaven National
  Lab.},\ \bibinfo {year} {1988})\BibitemShut {NoStop}%
\bibitem [{\citenamefont {Sakai}, \citenamefont {B{\"o}ttcher},\ and\
  \citenamefont {Schmidt}(1982)}]{sakai1982}%
  \BibitemOpen
  \bibfield  {author} {\bibinfo {author} {\bibfnamefont {Y.}~\bibnamefont
  {Sakai}}, \bibinfo {author} {\bibfnamefont {E.-H.}\ \bibnamefont
  {B{\"o}ttcher}}, \ and\ \bibinfo {author} {\bibfnamefont {W.}~\bibnamefont
  {Schmidt}},\ }\bibfield  {title} {\enquote {\bibinfo {title} {On the electron
  drift velocity in solid neon},}\ }\href@noop {} {\bibfield  {journal}
  {\bibinfo  {journal} {Zeitschrift f{\"u}r Naturforschung A}\ }\textbf
  {\bibinfo {volume} {37}},\ \bibinfo {pages} {87--90} (\bibinfo {year}
  {1982})}\BibitemShut {NoStop}%
\end{thebibliography}%

\end{document}